\documentclass[article]{aastex701}

\usepackage{graphicx}
\usepackage[version=4]{mhchem}

\newcommand{\nautilus}{Nautilus}

\begin{document}

\title{A Scalable Path to Astrometric Exomoon Discoveries with the Nautilus Space Observatory}

\author[orcid=0000-0002-4309-6343,sname='Wagner']{Kevin Wagner}
\affiliation{Steward Observatory, University of Arizona, Tucson, AZ, USA}
\email[show]{kevinwagner@arizona.edu}

\author[orcid=0009-0005-7181-9747,sname='Seung']{Sumin Seung}
\affiliation{Steward Observatory, University of Arizona, Tucson, AZ, USA}
\affiliation{Department of Earth Science Education, Seoul National University, Seoul, Republic of Korea}
\email{ssm248@snu.ac.kr}

\author[orcid=0000-0003-3714-5855]{D\'aniel Apai}
\affiliation{Steward Observatory, The University of Arizona, Tucson, AZ, USA}
\affiliation{Lunar and Planetary Laboratory, The University of Arizona, Tucson, AZ, USA}
\affiliation{Department of Earth, Atmospheric, and Planetary Sciences, Massachusetts Institute of Technology, Cambridge, MA, USA}
\email[]{apai@arizona.edu} 

\author[orcid=0000-0002-6137-0342,sname='Biancalani']{Enrico Biancalani}
\affil{Department of Astronomy, University of Maryland, College Park, MD, USA}
\affil{NASA Goddard Space Flight Center, Greenbelt, MD, USA}
\affil{Center for Research and Exploration in Space Science and Technology II (CRESST II), Greenbelt, MD, USA}
\email{ebiancalani94@gmail.com}

\author[orcid=0000-0002-9408-8925,sname='Bendek']{Eduardo Bendek}
\affiliation{NASA Ames Research Center, Moffett Field, CA, USA}
\email{eduardo.bendek@nasa.gov}

\author[orcid=0000-0002-4894-193X,sname='Hasler']{Samantha Hasler}
\affiliation{Department of Earth, Atmospheric, and Planetary Sciences, Massachusetts Institute of Technology, Cambridge, MA, USA}
\email{shasler@mit.edu}

\author[orcid=0000-0002-6087-3271,sname='Kostogryz']{Nadiia Kostogryz}
\affiliation{Max Planck Institute for Solar System Research, G\"ottingen, Germany}
\email{kostogryz@mps.mpg.de}

\author[orcid=0000-0002-3243-1230,sname='Krishnamurthy']{Sowmya Krishnamurthy}
\affiliation{University of Graz, Graz, Austria}
\email{sowmya.krishnamurthy@uni-graz.at}

\author[orcid=0000-0003-3204-8183]{Mercedes López-Morales}
\affiliation{Space Telescope Science Institute, 3700 San Martin Drive, Baltimore MD 21218, USA}
\email{mlopez-morales@stsci.edu}

\author[orcid=0000-0002-1052-6749,sname='McGill']{Peter McGill}
\affiliation{Lawrence Livermore National Laboratory, Livermore, CA, USA}
\email{mcgill5@llnl.gov}

\author[orcid=0000-0002-8864-1667,sname='Plavchan']{Peter Plavchan}
\affiliation{Department of Physics and Astronomy, George Mason University, Fairfax, VA, USA}
\email{pplavcha@gmu.edu}

\author[orcid=0000-0002-3627-1676,sname='Rackham']{Benjamin V. Rackham}
\affiliation{Department of Earth, Atmospheric and Planetary Sciences, and Kavli Institute for Astrophysics and Space Research, Massachusetts Institute of Technology, Cambridge, MA, USA}
\email{brackham@mit.edu}

\author[orcid=0000-0002-8842-5403,sname='Shapiro']{Alexander Shapiro}
\affiliation{University of Graz, Graz, Austria}
\email{alexander.shapiro@uni-graz.at}

\author[orcid=0000-0003-3989-5545,sname='Tuchow']{Noah Tuchow}
\affiliation{Steward Observatory, University of Arizona, Tucson, AZ, USA}
\email{nwtuchow@arizona.edu}

\author[orcid=0000-0001-8834-9108,sname='Worden']{S. Pete Worden}
\affiliation{The Breakthrough Prize Foundation, CA, USA}
\affiliation{Steward Observatory, University of Arizona, Tucson, AZ, USA}
\email{spworden@arizona.edu}

\author[orcid=0000-0003-2969-6040,sname='Zhou']{Yifan Zhou}
\affiliation{Department of Astronomy, University of Virginia, Charlottesville, VA, USA}
\email{yzhou@virginia.edu}


\begin{abstract}




Moons orbiting exoplanets (exomoons) can be detected through the reflex motion they impart to their host planet, which is recoverable in relative star--planet astrometric time series. The signal grows with moon mass and orbital separation and decreases with distance, so the nearest and least massive imaged planets are the most favorable targets. Recovering small ($\lesssim$\,Earth-mass) moons requires continuous, long-baseline, high-precision monitoring that is only practical with a dedicated or nearly dedicated facility. Building on recent simulations of astrometric exomoon detection and of the resulting population yields, we argue that the scalable, replicable architecture of the Nautilus Space Observatory is uniquely suited to this problem, and we outline a staged campaign. In an initial phase, one or a few small apertures target the nearest imaged giant planets---a high-reward but low-probability search focused on the closest stars. As the array is built out, the astrometric noise floor decreases and the same technique extends the search to the nearest such systems among nearby stars of spectral type K and earlier. This would be performed in parallel with high-contrast imaging and spectral characterization of the host planets and in synergy with a companion starshade concept for imaging Earth-like planets around the same nearby stars. Nautilus thus provides a scalable path from the first detection of a nearby exomoon toward a systematic search for exomoons around the closest stars.

\end{abstract}



\section{Nautilus Space Observatory White Paper}

This White Paper presents a potential science case for the Nautilus Space Observatory, a concept under development for a NASA Strategic Mission for the Astro2030 Decadal Survey. Nautilus is a constellation of space telescopes and will provide a modular, scalable, sustainable, upgradable, expandable space observatory that can be deployed rapidly and then expanded progressively. The core concept for Nautilus is described in \citet{apai2019}. This White Paper is part of the first series of science white papers capturing ideas that emerged from the Nautilus Science Case workshop (held at MIT in May 2026).

\section{Scientific Context and Problem Statement} 

Nearly every planet in the Solar System hosts moons, and the giant planets host rich satellite systems spanning a wide range of moon-to-planet mass ratios ($q$), from the Galilean satellites ($q\sim5\times10^{-5}$) through the Earth--Moon ($q\simeq0.012$) and Pluto--Charon ($q\simeq0.1$) systems. These satellites form through several channels, including accretion in a circumplanetary disk \citep{canup2006,cilibrasi2021}, giant impacts \citep{barr2017}, and gravitational capture \citep{agnor2006}. Several, including Europa and Enceladus, are themselves prime astrobiological targets owing to their subsurface oceans \citep{porco2006,hand2007}. By analogy, we expect moons orbiting exoplanets (exomoons) to be common, and they are compelling objects for studies of planetary system formation, dynamical evolution, and habitability. Those orbiting nearby habitable-zone planets are of particular astrobiological interest, since such moons could themselves host habitable environments \citep{reynolds1987,williams1997,scharf2006}.

No exomoon has been confirmed to date, despite searches using a range of techniques: transit-timing variations \citep{simon2007,kipping2009}, transits of the moons themselves \citep{teachey2018,kipping2022}, microlensing \citep{bennett2014}, spectroastrometry \citep{agol2015}, and direct imaging of tidally heated moons \citep{peterslimbach2013}. The most prominent candidates come from transit data, but their interpretations remain debated \citep{heller2019,kreidberg2019,hellerhippke2024}, and transit methods are biased toward more distant systems. Relative astrometry offers a complementary route that is strongest precisely for the nearest systems: an unseen moon displaces its host planet about the planet--moon barycenter, producing a periodic wobble in the measured star--planet separation with semi-amplitude $\propto M_{\rm moon}\,a_{\rm moon}/[(M_{\rm planet}+M_{\rm moon})\,d]$ \citep{lazzoni2022,wagner2025}; see also recent VLTI/GRAVITY interferometric programs that have placed analogous astrometric constraints on moons and binary companions of imaged substellar objects \citep{winterhalder2026method,kral2026,winterhalder2026betapic}. The same approach has placed companion limits in nearby binaries with HST \citep{bond2015} and motivates dedicated astrometric concepts such as TOLIMAN \citep{tuthill2018}. The nearest imaged giant planets are therefore the most favorable targets. The closest, $\alpha$\,Centauri\,A, hosts a habitable-zone giant-planet candidate indicated by VLT and JWST imaging \citep{wagner2021,beichman2025,sanghi2025}, with a mass below $\sim$100\,$M_\oplus$---comparable to Saturn \citep{zhao2018}. Moons there would be dynamically stable against the perturbation of $\alpha$\,Cen\,B \citep{rosariofranco2020,quarles2021}. \citet{wagner2025} showed that, at $0.1$\,mas precision with $1$\,hr cadence over a five-year campaign, moons down to $\sim$0.2\,$M_\oplus$ could be recovered around such a planet, and \citet{seung2026} extended this to general sensitivity scaling relations and population-integrated detection probabilities.

\textbf{Problem statement:} Confirming and characterizing exomoons---and ultimately measuring their occurrence---requires high-precision relative astrometry sustained continuously over multi-year baselines, with enough sensitivity to reach the low-mass moons that dominate plausible populations. Continuous, dedicated monitoring of individual targets over such baselines is impractical under the community-shared, multi-program model of current flagship observatories (e.g., HST and JWST), and the astrometric noise floor of any single aperture caps the accessible moon-mass range. Both limitations are addressed by a scalable array such as Nautilus \citep{apai2019,apai2022}, which can dedicate units to long-baseline monitoring while driving down the noise floor as it is built out. The same need for dedicated, long-duration staring on a small number of nearby targets also arises in starshade-based searches for Earth-like planets around the closest stars \citep{seager_prep}, motivating a shared or staged observatory architecture.

\section{Science Objectives}\label{sec:objectives}

We propose a program defined by three objectives (O1--O3), whose science scope scales with the observatory:

\begin{itemize}
\item \textbf{O1 --- Closest nearby targets (high reward, low probability).} With a single small aperture (a near-term, purpose-built $4$--$8.5$\,m telescope; e.g., \citealt{douglas2023}), search the nearest habitable-zone giant planet(s)---foremost the $\alpha$\,Cen\,A\,b candidate---for $\sim$Earth-mass and smaller moons. Other very nearby planet hosts---$\epsilon$\,Eridani \citep{hatzes2000,mawet2019}, $\epsilon$\,Indi\,A \citep{feng2019,matthews2024}, and $\tau$\,Ceti \citep{tuomi2013,feng2017}---extend the initial target list. Even a non-detection places astrophysically meaningful limits on moons around the nearest such worlds.
\item \textbf{O2 --- Nearest-system survey (general yield).} As identical units are added into an array, push the per-epoch astrometric precision from $\sim$0.1\,mas (single aperture) into the microarcsecond regime ($\lesssim$10--50\,$\mu$as) and extend the search to a volume-limited sample of nearby stars of spectral type K and earlier hosting imaged giant or sub-Neptune planets, beginning to map the exomoon mass--separation distribution rather than characterizing a single system. In the contrast-limited regime the accessible stellar sample grows from the few nearest systems with a single unit to $\sim$hundreds within $\sim$20\,pc for a $\sim$100-unit array, contingent on holding the systematic floor (per unit and pointing) to tens of $\mu$as (Section~\ref{sec:analysis} and Fig.~\ref{fig:yield}).
\item \textbf{O3 --- Formation and habitability.} Use the resulting detections and limits to constrain the moon-to-planet mass function, discriminate among satellite-formation channels \citep{canup2006,cilibrasi2021}, and identify candidate habitable moons. Because the moon itself is never detected---only the host planet's reflex---this characterization is dynamical: the recovered mass and orbit set the tidal-heating rate, flagging Europa- or Enceladus-like candidates for subsurface oceans. Direct characterization of a moon's surface or atmosphere lies beyond reflex astrometry, requiring a far fainter companion to be resolved from its host planet at sub-mas separations or identified through spectroastrometry \citep{agol2015}.
\end{itemize}

We stage the proposed program according to detectability. For the fiducial $\alpha$\,Cen configuration, the population-integrated probability of detecting at least the dominant moon of the giant planet is only of order one percent at $0.1$\,mas precision \citep{seung2026}. The detection probability rises by factors of several as the precision improves to $10$--$50$\,$\mu$as, and the detectable fraction of plausible moons climbs steeply once the mass sensitivity reaches $\lesssim$0.05\,$M_\oplus$ \citep{seung2026}. The single-aperture phase is thus a low-probability/high-reward search of the nearest systems, while placing meaningful constraints on nearby planets and moons. The array extends the same technique to more distant systems$-$enabling population-level studies.

\section{Data Requirements}\label{sec:datareq}

The core measurement is precise \emph{relative} astrometry of an imaged planet with respect to its host star, in reflected light. Reflected-light flux peaks in the optical, so the baseline astrometric band is $\sim$500\,nm, with broad optical (B/G/R) coverage desirable for color and characterization. The raw contrast required to detect and centroid a reflected-light giant is $\sim$10$^{-9}$ at the relevant separations, a level that recent laboratory results suggest is achievable (\citealt{anche2024}; Table~\ref{scireq}). The defining requirements are temporal: a per-epoch relative-astrometric precision of $\sim$0.1\,mas for the single-aperture phase, improving to $\lesssim$10--50\,$\mu$as for the full array; an observing cadence of order one hour; and a continuous baseline of $\gtrsim$5\,yr (extending toward a decade), set by the need to phase-fold the moon signal while sampling the planet's orbit \citep{wagner2025,seung2026}. Sensitivity improves approximately linearly with astrometric precision and only weakly with the number of epochs, $N_{\rm obs}$, set by the cadence and total duration ($\propto N_{\rm obs}^{-1/2}$), so precision---and hence collecting area and stability---is the primary driver \citep{seung2026}. We emphasize that hourly cadence and strictly continuous coverage are advantageous rather than fundamental: detection is limited primarily by the per-epoch precision, and the moon-mass sensitivity degrades only gradually with coarser sampling, following the approximate $N_{\rm obs}^{-1/2}$ scaling found by \citet{seung2026}---for example, relaxing the cadence from 1 to 10\,hr costs a factor of $\sim$3 in minimum detectable moon mass. The firm sampling requirement is only that the cadence resolve the $\sim$4--30\,d moon periods---even daily sampling remains viable for larger moons \citep{wagner2025}---so the principal cost of sparser observations is the reduced number of epochs rather than aliasing. Systematic terms, such as determining the star's position behind the coronagraph and calibrating field distortion, must be controlled to a comparable level; these are currently of order $\sim$mas \citep{maire2021,weible2025} and represent the central calibration challenge. Astrophysical noise sources, by contrast, are subdominant or mitigable: the astrometric jitter of nearby Sun-like stars from starspots and faculae is at the $\sim\mu$as level---below the per-epoch precisions adopted here and further reducible through multiwavelength and temporal modeling \citep{kaplanlipkin2022,deagan2026}---as are photocenter shifts from planetary ring shadows and orbital smearing \citep{wagner2025}. These requirements are summarized in Tables~\ref{scireq} and \ref{scireq2}.

\section{Analysis and Interpretation}\label{sec:analysis}

\begin{figure*}[ht!]
\centering
\includegraphics[width=\textwidth]{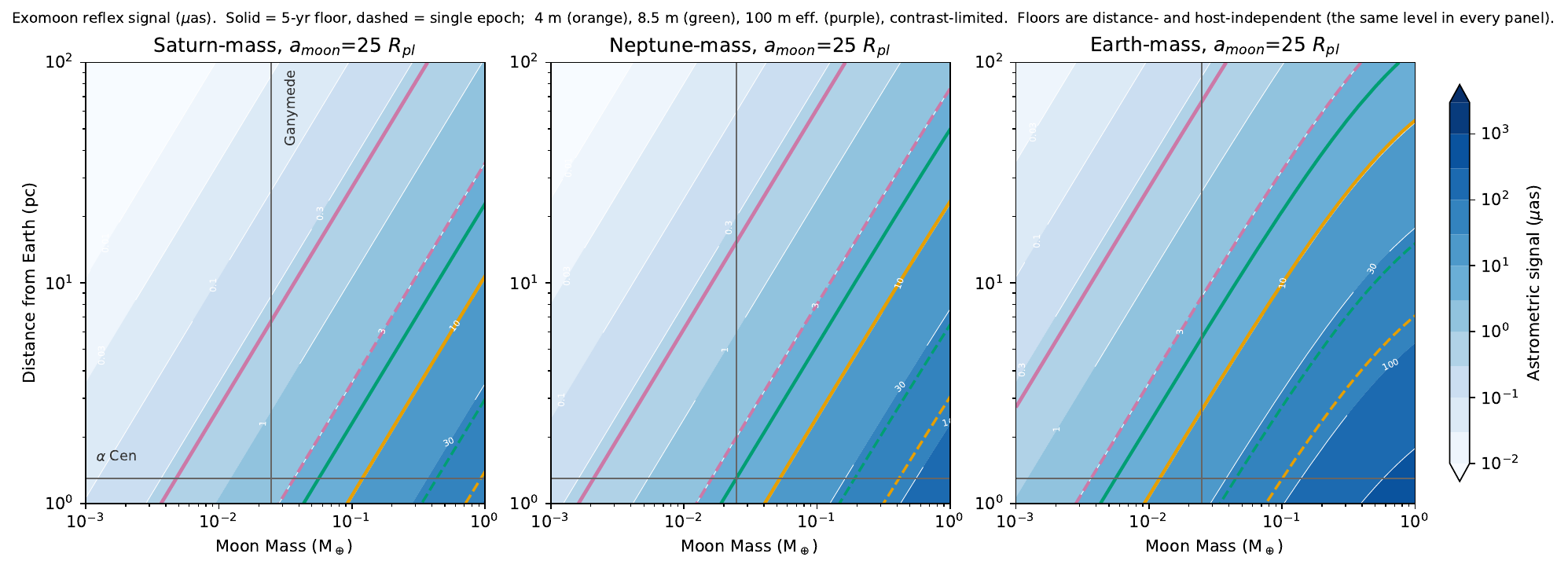}
\caption{Astrometric reflex signal (color scale, $\mu$as) induced by an exomoon as a function of moon mass and system distance, for Saturn-, Neptune-, and Earth-mass host planets at a moon semi-major axis $a=25\,R_{\rm pl}$. Colored curves mark contrast-limited detection limits for three facility scales---$4$\,m (orange), a single $8.5$\,m unit (green), and a $100$\,m effective collecting diameter (purple)---with solid lines the five-year campaign floor and dashed lines the single-epoch precision. As described in Section~\ref{sec:analysis}, the contrast-limited floor is a fixed signal level, identical across the three panels, which differ only through the host's reflex amplitude. Thin gray lines mark the Ganymede-mass moon ($0.025\,M_\oplus$) and the distance of $\alpha$\,Centauri ($1.3$\,pc).}
\label{fig:siggrid}
\end{figure*}

The astrometric signal of an exomoon is the reflex motion of the host planet about the planet--moon barycenter, with semi-amplitude $\theta = [M_{\rm moon}/(M_{\rm planet}+M_{\rm moon})]\,(a_{\rm moon}/d)$, where $M_{\rm moon}$ and $M_{\rm planet}$ are the moon and planet masses, $a_{\rm moon}$ is the moon's orbital semi-major axis, and $d$ is the system distance. This signal is recovered by fitting and removing the planet's orbit about the host star from the relative star--planet astrometric time series and searching the residuals for the periodic, moon-induced wobble \citep{wagner2025,winterhalder2026method}. Figure~\ref{fig:siggrid} maps this signal across moon mass and system distance for representative host planets (Saturn-, Neptune-, and Earth-mass) at a fixed moon semi-major axis of $25\,R_{\rm pl}$, overlaid with the single-epoch and five-year-campaign detection limits for several facility scales. Following \citet{wagner2025} and \citet{seung2026}, we adopt a five-year-campaign detection floor of $0.13\times$ the per-epoch precision---the value that reproduces their $\sim$0.2\,$M_\oplus$ limit at $\alpha$\,Cen. In their simulations, a moon is counted as recovered when a sinusoidal fit to the phase-folded residuals improves the reduced $\chi^2$ over a flat (no-moon) model by more than $5$ and returns the moon's period and barycentric semi-amplitude to within $5\%$ and $25\%$ of their true values, respectively. The detection fraction is the share of Monte Carlo trials satisfying all three criteria. The recovered semi-amplitude in turn yields the moon-to-planet mass ratio. 

Because a lighter host exhibits a larger reflex motion for a given moon, its moons are detectable to larger distances: for an Earth-mass host, Ganymede-scale moons ($0.025\,M_\oplus$) clear the detection floor around the nearest systems even with a single $8.5$\,m unit, whereas a Saturn-mass host requires either a moon a few times more massive or a larger effective aperture. In adopting this case we also inherit the orbital assumptions of \citet{wagner2025}: Keplerian orbits for both the planet about the star and the planet--moon pair about their barycenter, a single dominant moon, and a planet whose year-scale orbital period is fully sampled within the campaign---and therefore well separated in frequency from the $\sim$4--30\,d moon signal, so that the planet's reflex can be fit and removed with minimal covariance with the moon term. These idealizations are expected to have only modest impact: the moon's orbital inclination and phase enter as an order-unity geometric projection factor on the recovered amplitude, while orbital eccentricity or additional moons redistribute signal power across harmonics without changing the characteristic threshold mass.

\begin{figure*}[ht!]
\centering
\includegraphics[width=\textwidth]{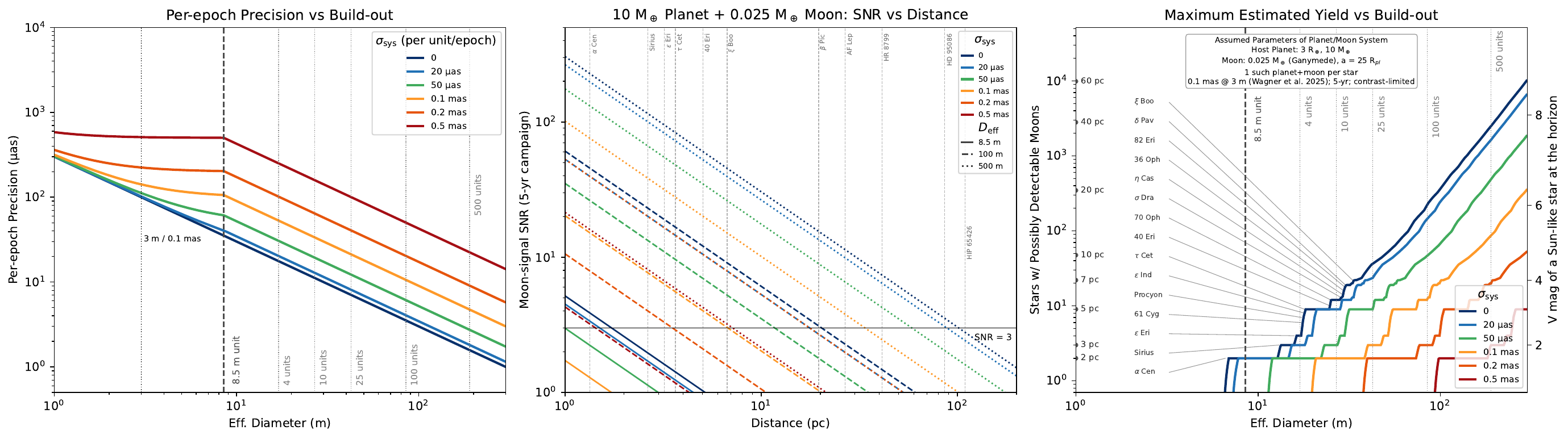}
\caption{Contrast-limited astrometric exomoon yield for a \nautilus-like build-out (a single mirror up to $8.5$\,m, then $N$ identical $8.5$\,m units, $D_{\rm eff}=8.5\sqrt{N}$\,m, to a $\sim$300\,m-effective array), shown for various systematic per-unit/per-epoch astrometric floors $\sigma_{\rm sys}=0$--$0.5$\,mas (color-coded); the fiducial system, anchoring, and yield scalings are detailed in Section~\ref{sec:analysis}. \textit{Left:} per-epoch precision versus build-out---a single mirror saturates at $\sigma_{\rm sys}$, whereas the array keeps improving as $\sigma\propto D_{\rm eff}^{-1}$. \textit{Center:} moon-signal SNR versus distance for three effective collecting diameters ($8.5$, $100$, and $500$\,m; line style); each curve's SNR\,$=3$ crossing is its detection horizon. Vertical dashed lines mark nearby K-type and earlier stars and several young, directly-imaged planetary systems ($\beta$\,Pic, AF\,Lep, HR\,8799, HD\,95086, HIP\,65426). \textit{Right:} maximum estimated number of K-type and earlier stars (M dwarfs excluded) hosting a detectable moon; left-hand markers give the survey-distance horizon (pc), the right axis the apparent $V$ magnitude of a Sun-like host there, and laddered names the individual nearby systems entering the sample.}
\label{fig:yield}
\end{figure*}

To assess the observatory requirements, we model the per-epoch astrometric precision as a function of effective collecting aperture, anchored at $0.1$\,mas for a $3$\,m telescope (consistent with a contrast-limited performance of $\sim$$10^{-9}$ at $500$\,nm). Up to the $8.5$\,m single-telescope size, enlarging the primary sharpens the diffraction-limited point spread function, so the contrast-limited per-epoch precision, $\sigma$, improves with the telescope diameter, $D$, as $\sigma\propto D^{-1}$; beyond this size, additional identical $8.5$\,m units drive $\sigma\propto N_{\rm units}^{-1/2}$, which coincides with $\sigma\propto D_{\rm eff}^{-1}$ in effective diameter. As a demonstration of the possibly accessible sample size, we adopt a fiducial system---a sub-Neptune-mass planet ($10\,M_\oplus$, radius $\sim$3\,$R_\oplus$) hosting a Ganymede-mass moon ($0.025\,M_\oplus$) at $a=25\,R_{\rm pl}$---and assess its detectability over a five-year campaign under various assumptions on the per-epoch and final astrometric precision. We count K-type and earlier nearby stars from the Gaia Catalogue of Nearby Stars \citep{gcns2021}, excluding M dwarfs (whose reflected-light planets are too faint for this measurement) and supplementing the brightest, Gaia-saturated systems within $7$\,pc from a curated list. Figure~\ref{fig:yield} shows the per-epoch precision (left), the moon-signal SNR versus distance for effective diameters of $8.5$--$500$\,m (center), and the maximum estimated number of stars with detectable moons versus build-out (right), each plotted for a set of systematic floors $\sigma_{\rm sys}$ (below). The precision is anchored to $\alpha$\,Cen\,Ab---a Saturn-size giant planet candidate at $1.3$\,pc---and the $10\,M_\oplus$ host enters through the moon-to-planet mass ratio scaling, which sets the reflex amplitude. The count in Figure~\ref{fig:yield} should be interpreted as an approximate upper bound on the number of systems with detectable moons, since it assumes that every such star hosts the fiducial planet--moon system. It does not account for the possibility of multiple planets within a given system or multiple moons per planet. Thus, the actual yield of exomoons could be even larger should the average number of detectable moons per system be $>$1. Several young, directly-imaged systems ($\beta$\,Pic, AF\,Lep, HR\,8799, HD\,95086, HIP\,65426) are also marked in Figure~\ref{fig:yield} for context. These planets are self-luminous and bright in thermal emission, but on wide orbits and therefore faint in reflected light. Detecting moons around these known planets would require coverage in the infrared (Tables~\ref{scireq} and \ref{scireq2}).

We assume the per-epoch and per-unit measurements (coronagraphic imaging or imaging with a starshade: \citealt{seager_prep}) are contrast-limited: i.e., the astrometric noise is residual starlight (speckles) rather than planet photon counts. Because that stellar noise falls as $1/d^{2}$ in lockstep with the reflected planet signal, the per-epoch SNR---and hence the precision---is independent of distance, and the recoverable moon signal-to-noise declines only as $1/d$ (Figure~\ref{fig:yield}, center). The detection horizon therefore grows in direct proportion to the precision gain---until it is capped by a systematic astrometric floor, $\sigma_{\rm sys}$: a per-unit, per-epoch error (e.g., determining the star's position behind the coronagraph or starshade) that is independent of aperture and statistically independent between units and epochs, so it adds in quadrature to the per-epoch precision and averages down as $N_{\rm units}^{-1/2}$ over the array and as $N_{\rm obs}^{-1/2}$ over the campaign. Figure~\ref{fig:yield} maps the yield for $\sigma_{\rm sys}=0$--$0.5$\,mas. A floor below the single-unit precision ($\sim$35\,$\mu$as for an $8.5$\,m unit) is negligible---$\sigma_{\rm sys}\lesssim20\,\mu$as is indistinguishable from zero---whereas $\sigma_{\rm sys}\sim50$--$100\,\mu$as limits even a $\sim$300\,m-effective array to $\sim$20--35\,pc, and the systematic level of present high-contrast astrometry ($\sim$mas; \citealt{maire2021,weible2025}) would collapse the reach to $\sim$2\,pc and a handful of the nearest systems. A genuine population census ($\sim$$10^{2}$--$10^{3}$ systems within tens of parsecs at the $\sim$100--300\,m scale) therefore requires pushing the systematic floor to tens of $\mu$as---the central technical challenge. Because $\sigma_{\rm sys}$ is independent between units, a replicable array averages it down by $\sqrt{N_{\rm units}}$, while a single monolithic aperture of equal collecting area remains pinned at the per-epoch floor---a concrete advantage of the replicable Nautilus architecture. We treat $\sigma_{\rm sys}$ as fully random; any component correlated between epochs would not average over the campaign and would act as an irreducible floor, making its control still more critical. A pure planet-shot-noise limit, in which precision instead degrades with distance, is far more pessimistic but is not the operative regime for these high-contrast, speckle-limited measurements.

\begin{figure*}[ht!]
\centering
\includegraphics[width=\textwidth]{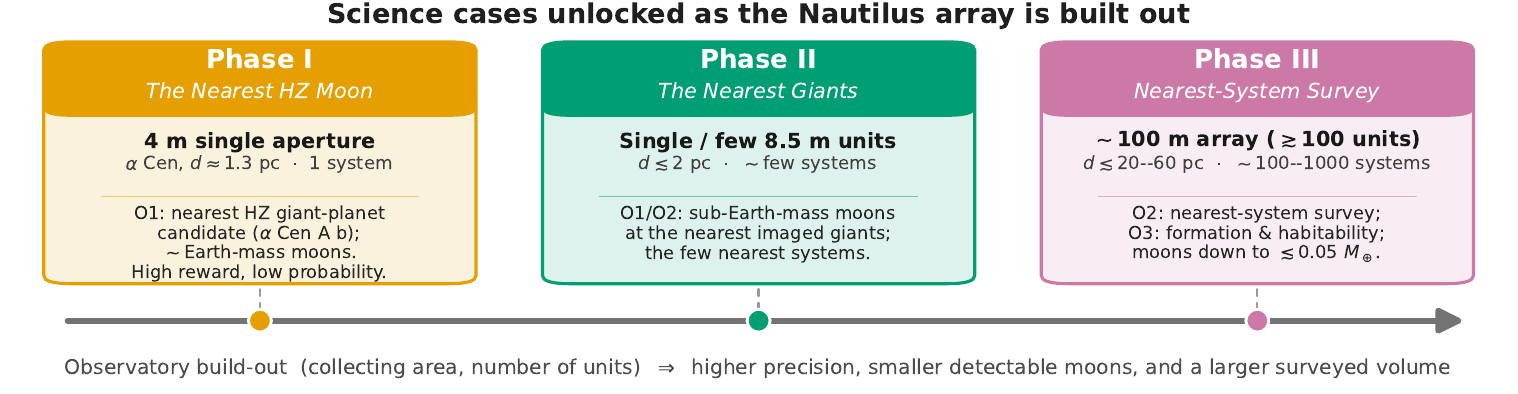}
\caption{Science cases unlocked as the Nautilus array is built out. Each phase lists the configuration, the approximate survey horizon and accessible stellar sample, and the associated objectives (O1--O3), as developed in Sections~\ref{sec:datareq}--\ref{sec:analysis}.}
\label{fig:timeline}
\end{figure*}

These detection limits translate into a population-integrated yield only when folded through an assumed exomoon distribution, analogous to exoplanet-yield estimates for direct-imaging missions \citep{quanz2022}. \citet{seung2026} carried out this calculation for the $\alpha$\,Cen giant-planet candidate, combining the astrometric success maps with empirical (Solar System plus Kepler-candidate) and population-synthesis \citep{cilibrasi2021} moon distributions, taking each system's signal to be set by its single largest moon. They find a probability of recovering the dominant moon of order $\sim$0.5--3\% at $0.1$\,mas precision, rising to several percent at $10$--$50$\,$\mu$as, with the detectable fraction of moons increasing steeply once the mass sensitivity falls below $\sim$0.05\,$M_\oplus$. This is the quantitative basis for the staged strategy summarized in Figure~\ref{fig:timeline}: a single small aperture makes the nearest habitable-zone system a high-reward, low-probability target, whereas the array's improved noise floor and larger accessible volume extend the method to a population of nearby systems---of order $10^{2}$ within $\sim$20\,pc for a $\sim$100-unit array, growing toward $\sim$$10^{3}$ within tens of parsecs at the $\sim$300\,m scale, contingent on holding the systematic floor to tens of $\mu$as (Fig.~\ref{fig:yield}). Across that sample, the program begins to constrain the occurrence rate and population-level trends of exomoons.

\section{Relevant Science Requirements}

The principal data and system-level requirements implied by the science case are summarized in Tables~\ref{scireq} (spectral coverage) and \ref{scireq2} (observatory and observing parameters), as first-order estimates to be refined as the Nautilus trade space matures (see other white papers in this series). The core measurement is made in reflected light---broad optical imaging, with near-infrared coverage optional for characterization---of a narrow ($\sim$few-arcsec) field near the diffraction limit ($\leq$5\,mas) at a raw contrast of $\lesssim$$10^{-9}$. The defining requirement is a relative astrometric precision of $\lesssim$0.1\,mas per epoch (improving to $\lesssim$10--50\,$\mu$as for the full array), sufficient to detect $\sim$0.2--1\,$M_\oplus$ moons. The driving needs are temporal---e.g., $\sim$1\,hr exposures and cadence over a continuous $\sim$5\,yr baseline. These requirements of deep contrast and long, repeated integrations on a small list of nearby targets closely parallel those of the proposed starshade reflected-light imaging concept \citep{seager_prep}, pointing to shared hardware and observing strategies.

\begin{deluxetable}{lccl}
\tabletypesize{\scriptsize}
\tablecaption{Wavelength-coverage requirements for imaging and spectroscopy of the host planets. \label{scireq}}
\tablewidth{0pt}
\tablehead{
\colhead{Requirement} & \colhead{Imaging} & \colhead{Spect.} & \colhead{Science Driver} 
}
\startdata
250--350\,nm  &  N/A &  N/A & Limited reflected-light flux\\
350--450\,nm  & Required & Optional & B-band reflected light \\
450--1,000\,nm  & Required & Optional & G/R-band reflected light; 500\,nm astrometric reference (Wagner et al. 2025)\\
1--1.8\,$\mu$m  & Optional & Optional & Planet thermal emission / characterization \\
1.8--2.3\,$\mu$m  & Optional & Optional & Young planets \\
2.3--2.9\,$\mu$m  & Optional &  Optional & Young planets\\
\enddata
\end{deluxetable}

\begin{deluxetable}{lll}
\tabletypesize{\scriptsize}
\tablecaption{Observatory and observing-campaign requirements for astrometric exomoon detection. \label{scireq2}}
\tablewidth{0pt}
\tablehead{
\colhead{Requirement} & \colhead{Range} & \colhead{Science Driver} 
}
\startdata
Photometric Filters & BVRI & Reflected light from HZ planets\\
 & JHK & Young giant planets\\
Target Brightness [mag] & 0--10 & Nearby stars and young stars  \\
Min. Photom. Precision [ppm] & N/A & Signal is astrometric, not photometric \\
Image Res. [diff. limit] & $\leq$5 mas & Astrometric prec. $\lesssim$0.1 mas w/ SNR$\gtrsim$50 \\
Astrometric Precision & $\lesssim$0.1 mas/epoch & Detect $\sim$0.2--1 $M_\oplus$ moons (Wagner et al. 2025) \\
 & $\lesssim$10--50 $\mu$as (array) & Smaller moons \& larger volume (Fig.~\ref{fig:yield})\\
Min. Sky Coverage [arcsec] & $\sim$few arcsec&  Narrow FoV per target\\
Min. Contrast & $\lesssim$10$^{-9}$ & Reflected-light HZ giant (1\,hr, 500\,nm); w.r.t. $\alpha$ Cen A at $1''$ (Wagner et al. 2025)\\
Spectral Resolving Power & N/A & Broadband astrometry; spectroscopy optional \\
Telescope Aperture & $\geq$4\,m, or $N\times$8.5\,m units & Precision \& yield scaling (Fig.~\ref{fig:yield})\\
\hline
Relevant Timescales [s] & 3600 & 1\,hr exposure per epoch (Wagner et al. 2025) \\
Monitoring Baseline [d] & $\sim$1825 (5\,yr) & Sample planet orbit \& moon periods (Wagner et al. 2025) \\
Cadence [s] & 3600 & $\leq$1\,hr; resolve 4--30\,d moon periods (Wagner et al. 2025) \\
Rapid Response Time [s] & N/A & Long-baseline monitoring; not time-critical \\
\hline
Data Volume & High  & Continuous narrow-field astrometric time series \\
Pointing Precision [mas] & $\lesssim$0.1 mas & Relative star--planet astrometry \\
\enddata
\end{deluxetable}

\section{Relevance to Nautilus and Mission Class}

Nautilus---a constellation of replicable units, each built around a large, ultralight multi-order diffractive element (MODE) lens \citep{apai2019,apai2022}---is well suited to this science for three reasons. First, and most distinctively, the measurement demands \emph{years of continuous, dedicated} monitoring of individual targets. This is fundamentally incompatible with the operating model of today's flagship facilities: HST and JWST are heavily oversubscribed, community-shared observatories that must serve hundreds of programs spanning the full breadth of astrophysics, and cannot commit an instrument to staring at a single star for years. A Nautilus unit, by contrast, can be dedicated to one target for the entire multi-year baseline the science requires; because the units are mass-produced from a common, scalable design \citep{apai2019,apai2022}, doing so does not starve the rest of the community---additional units simply expand the number of targets monitored in parallel. Nautilus therefore fills a capability gap that no current or planned flagship addresses: dedicated, long-baseline, high-cadence astrometric monitoring carried out at the scale of a survey. Second, sensitivity is set by the astrometric noise floor, which the array improves as units are added: in the contrast limit the per-epoch precision scales as $D^{-1}$ while a single mirror grows toward the $\sim$8.5\,m unit size, and then as $N_{\rm units}^{-1/2}$ as identical units are added (Fig.~\ref{fig:yield}), so the accessible moon-mass range---and hence the expected yield---grows directly with build-out. Because the units form an incoherent array rather than a phased interferometer, the diffraction-limited angular resolution and inner working angle remain those of the single $8.5$\,m unit and do not improve with build-out; what grows is the signal-to-noise on an already-imaged planet, exactly what relative astrometry---a centroiding rather than a resolving measurement---requires. The units could in principle be phased to synthesize finer angular resolution, but the reflex-astrometry measurement does not require it. Crucially, per-unit systematic errors that are independent between units (e.g.\ coronagraphic star-centering) also average down as $N_{\rm units}^{-1/2}$, so the replicable array suppresses the very calibration floor that limits the measurement, whereas a single monolithic aperture of equal collecting area cannot---turning the dominant systematic into a quantity that improves with build-out (Figure~\ref{fig:yield}). The replicable-unit design lets the same hardware perform high-contrast imaging and spectral characterization of the host planets in parallel, and observe many targets simultaneously once the array is built out.

Starlight suppression for the astrometric census is baselined as a coronagraph on each unit, since no single external occulter can shadow the full constellation at once---an occulter sized to cover a $\sim$100\,m-effective array of free-flying units would far exceed the $\sim$35\,m starshade needed for a $6$\,m aperture. A starshade is not thereby excluded for the census, however: because each target is monitored in long, sequential stares of months to years, the weeks-scale reslew of an occulter becomes a negligible overhead, so a single starshade could feed the array one target at a time, turning its usual slew-time penalty into a non-issue for this observing mode. This program is synergistic with a companion concept in this white-paper series \citep{seager_prep}: a starshade-equipped Nautilus unit (or units) conducting deep, multi-hour integrations on the nearest stars ($d\lesssim10$\,pc) to image and characterize Earth-like planets, building on the starshade direct-imaging concept developed in earlier mission studies such as Exo-S \citep{cash2006,seager2015exos,exos2015}. Such a campaign requires precisely the long, dedicated stares on individual nearby systems that exomoon astrometry also demands, and the starshade's deep starlight suppression aids detection of the reflected-light giant planets whose moons we seek. A small ($D\sim$4\,m telescope) Nautilus+starshade pathfinder could therefore advance both goals on the same targets---either in an early, shared configuration or as a precursor to the full array.

\textbf{Constellation:} The science benefits strongly from a constellation and from a multi-stage approach. An initial phase of one or a few units targets the uniquely positioned nearby systems (O1) and validates the technique, while a later, larger array delivers the nearest-system survey (O2--O3) by both lowering the noise floor and parallelizing across many targets. This maps naturally onto a phased deployment of identical units envisioned for Nautilus.

\textbf{Relevant Class:} The closest-targets phase is plausibly Probe to Flaglet class (one to a few $\sim$4--8.5\,m units), whereas the full nearest-system survey---which requires an effective collecting aperture assembled from many units---is Flagship class. The staged structure delivers compelling science, and meaningful risk reduction, at each investment level.

\section{Relevance to NASA and Astrophysics Strategy}

The search for life beyond the Solar System and the characterization of nearby worlds are central to the Astro2020 Decadal Survey \citep{astro2020}, which prioritized a future large UV/optical/IR capability and the broader goal of identifying and characterizing potentially habitable environments. Exomoons extend this goal: the icy and rocky moons of giant planets are themselves candidate habitats, and around the nearest stars they are accessible to the same reflected-light, high-contrast platforms being developed for exo-Earth science. A Nautilus exomoon campaign complements the Habitable Worlds Observatory \citep{feinberg2024} by targeting a distinct and largely unexplored population, advances the decadal priority of placing the Solar System's worlds---including its many moons---in a broader exoplanetary context, and matures the precise relative-astrometry and long-baseline-monitoring techniques that are broadly relevant across NASA's exoplanet program.

\begin{acknowledgments}
We thank the Heising-Simons Foundation for supporting the Nautilus Science Case Workshop. KW acknowledges support from The Breakthrough Prize Foundation. The results reported herein benefited from collaborations and/or information exchange within NASA's Nexus for Exoplanet System Science (NExSS) research coordination network sponsored by NASA's Science Mission Directorate. This research made use of NASA's Astrophysics Data System (ADS); the VizieR catalogue access tool and SIMBAD database operated at CDS, Strasbourg, France; and the Gaia Catalogue of Nearby Stars derived from data of the ESA \textit{Gaia} mission. PM acknowledges that this work was performed in part under the auspices of the U.S. Department of Energy by Lawrence Livermore National Laboratory under the contract DE-AC52-07NA27344, LLNL-CONF-2020815. 
\end{acknowledgments}

\bibliography{sample701,exomoon_refs}{}
\bibliographystyle{aasjournalv7}

\end{document}